# Fair and Square: Replacing One Real Multiplication with a Single Square and One Complex Multiplication with Three Squares When Performing Matrix Multiplication and Convolutions

Vincenzo Liguori            Ocean Logic Pty Ltd             enzo@ocean-logic.com

**Abstract**
**This paper shows that, for matrix multiplications and convolutions, it is possible to asymptotically replace each real multiplication with a single squaring operation. Similarly, a single complex multiplication can be replaced with 3 squaring operations. Given that an n bits squaring circuit requires about half the gate count of an nxn multiplier, this results in large resource reductions. With some caveats, the same techniques can apply to dot products, transformations and more. A varietiy of architectures implementing these ideas in hardware are described, including square based systolic arrays and tensor cores.**

# 1. Introduction

Operations such as matrix multiplications and convolutions are ubiquitous and essential in AI, DSP and many other applications. It follows that any optimisations in their implementation is extremely important.

This paper describes, for these operations, replacing (asymptotically), one real multiplication with one square operation and one complex multiplication with 4 and 3 squares. Given that an n bits squaring circuit requires about half the gate count of an nxn multiplier [1], this results a large resource reductions.

Section 2 describes the basic mechanism which is then applied in the subsequent sections, starting with matrix multiplications in section 3. This is followed, in sections 4 and 5, by the application to linear transforms, convolutions and correlations.

These sections also include the description of various hardware architectures implementing these ideas.

Sections 6, 7 and 8 extend these concepts to complex multiplications with 4 squares, 3 squares in sections 9, 10 and 11.

This paper transcends the particular implementation of a squaring circuit. Approximate squaring is also a possibility.

Note that many variable names and symbols are re-used in the paper. Their potentially different meanings should be obvious from the context. Normally the life of such objects is limited to a section or sub-section.

# 2. The Basic Mechanism

The basic idea in this paper is to replace multiplications with either of the following expression in the definition of various mathematical operations and see where that takes us:

$$(a+b)^2 = a^2 + b^2 + 2ab \Rightarrow ab = \frac{1}{2}((a+b)^2 - a^2 - b^2) \quad (1)$$

$$(a-b)^2 = a^2 + b^2 - 2ab \Rightarrow -ab = \frac{1}{2}((a-b)^2 - a^2 - b^2) \quad (2)$$

In each case we will see that all such operations can be redefined in terms of squaring operations.

## 3. Matrix Multiplication

Let us consider an MxN matrix A, an NxP matrix B, both real valued, and their product, the MxP matrix C with $a_{ik}$, $b_{kj}$ and $c_{ij}$ their respective elements:

$$c_{ij} = \sum_{k=0}^{N-1} a_{ik} b_{kj} \quad \forall i=0\ldots M-1, j=0\ldots P-1 \quad (3)$$

We can now use (1) to replace the multiplication $a_{ik}b_{kj}$:

$$c_{ij} = \sum_{k=0}^{N-1} a_{ik} b_{kj} = \frac{1}{2} \sum_{k=0}^{N-1} \left((a_{ik}+b_{kj})^2 - a_{ik}^2 - b_{kj}^2\right) = \frac{1}{2}\left(Sab_{ij} + Sa_i + Sb_j\right) \quad (4)$$

with:

$$Sab_{ij} = \sum_{k=0}^{N-1}(a_{ik}+b_{kj})^2, \quad Sa_i = -\sum_{k=0}^{N-1} a_{ik}^2, \quad Sb_j = -\sum_{k=0}^{N-1} b_{kj}^2 \quad (5)$$

This substitutions appears to make things worse, but only apparently: the terms $Sa_i$ depends on the i index only while $Sb_j$ depends on j only. They are simply the sum of the squares of the elements of the rows of A and the columns of B. They can be calculated separately, with M*N squares and N*P squares respectively, and reused for each $c_{ij}$.

The multiplication in (3) is replaced by the square of a sum in (4) plus the two additional terms. Effectively, the square of a sum can be seen as a "partial multiplication": an analog of the multiplication in (3).

To recap, in order to calculate all the elements of C with (3), we need M*N*P multiplications.

When we do the same with (4), we need M*N*P squaring operations plus, due to the re-use of the terms $Sa_i$ and $Sb_j$, M*N and N*P squaring operations respectively. Total M*N*P + M*N + N*P squaring operations.

Thus, calculating the ratio between the number of squaring operations and multiplications to calculate the matrix product, we obtain:

$$\frac{MNP + MN + NP}{MNP} = 1 + \frac{1}{P} + \frac{1}{M} \quad (6)$$

The ratio tends quickly to 1 as M and P grow. Essentially we can perform matrix multiplication with a single square operation per multiplication for any reasonable size matrices.

Often, in the case of AI inference, one of the two matrices, is to be considered constant and that means that either of the terms $Sa_j$ or $Sb_j$ can be pre-calculated.

## 3.1. Hardware Implementation Examples

Equations like (3) are normally implemented with multiply accumulators (MACs). Here we propose to replace the multiplier with a partial multiplier. Provided that we also include the additional terms in (4), the final result will be the same.

For example, Fig1a shows a MAC that is used to calculate (3). The register initialised to zero. Then pairs $a_{i0}$ and $b_{0j}$ are input. These are multiplied and accumulated in the register. These are followed by pair $a_{i1}$ and $b_{1j}$, $a_{i2}$ and $b_{2j}$, etc. until we input $a_{iN-1}$ and $b_{N-1j}$. After this the register will contain the value $c_{ij}$.

Fig1b shows a partial multiplication accumulator that works in a similar way to the MAC and can be used to calculate (4). We initialise its register first with $Sa_i + Sb_j$. We then input the pairs $a_{i0}$ and $b_{0j}$. These are added together, squared and accumulated in the register. We do the same for $a_{i1}$ and $b_{1j}$, etc, until input $a_{iN-1}$ and $b_{N-1j}$. After this its register will contain the value $2c_{ij}$, necessitating of a simple right shift to recover $c_{ij}$.

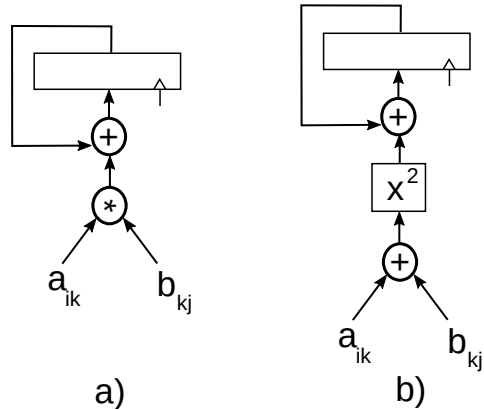

Figure 1: Multiply accumulator a). Partial multiplication accumulator b).

## 3.2. Square Based Systolic Arrays

The example above is very simple but it doesn't have to be. Fig.2 shows an example of stationary systolic array whose Processing Elements (PEs, Fig.3) have been modified by substituting the multiplier with a partial multiplier. The other modification allows to add the additional terms $Sa_i$ and $Sb_j$.

The operation of the systolic array is otherwise almost identical to a normal one (see [2]). First the values $a_{ij}$ are shifted into the REGA registers of each PE by selecting 0 in the PEs' mux (see Fig.3).

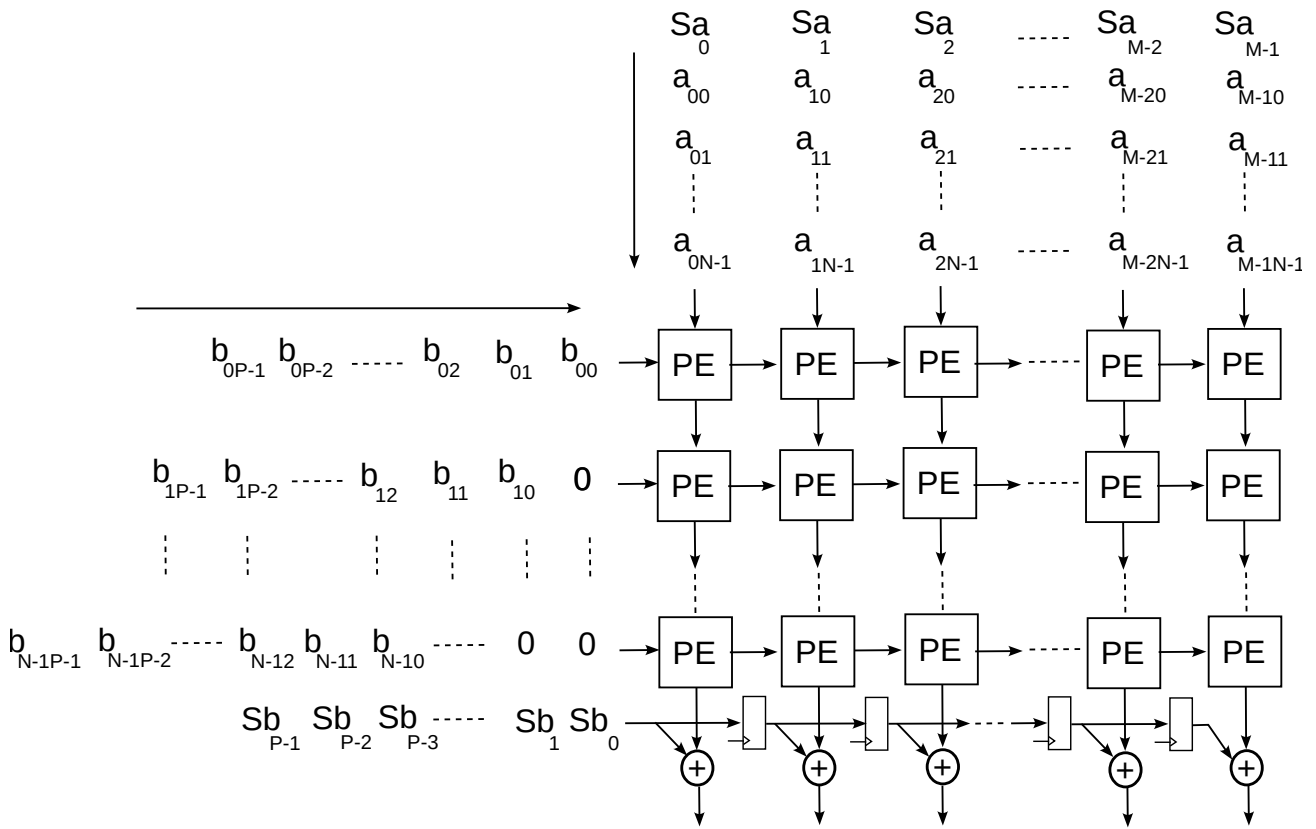

Figure 2: Example of square based systolic array.

Once the array has been loaded, the Sa$_i$ elements are now at the inputs at the top, providing the initial value of the subsequent sums. Then the elements b$_{ij}$ are shifted in, staggered.

As soon as the first result starts to emerge from the bottom left corner of the array, we start to shift in the elements Sb$_j$ which are added and finalise the results. Note that this design outputs the cij values multiplied by 2: we would need a simple shift right to obtain the result of the matrix multiplication.

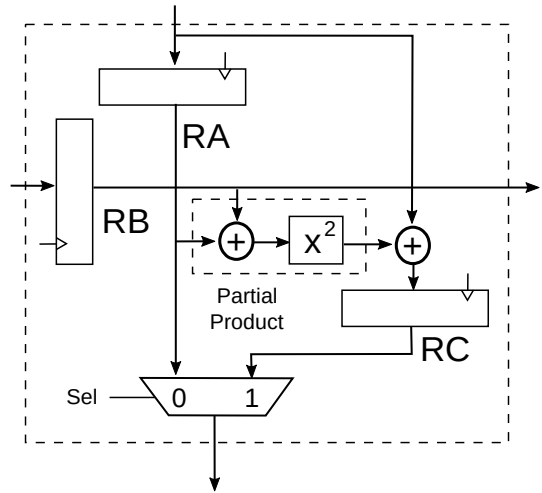

Figure 3: Processing Element (PE) for square based systolic array.

The architecture above is just an example: replacing the multiplier with a partial multiplier will work in any other systolic array architectures as long as we find a way to add the additional terms Sa$_i$ and Sb$_j$ to the final result.

A few words about Sa$_i$ and Sb$_j$. If the size of the systolic array is the same as the matrices being multiplied, then Sa$_i$ and Sb$_j$ could be calculated on the fly as the elements of A and B arrive at the array. This will save storage and/or multiple passes.

However, this is not very likely: normally the systolic array is smaller than the matrices being multiplied and the multiplication is done by tiling. This can complicate things and it might be simpler calculating the additional terms when the matrices they belong to are being created.

## 3.3. Square Based Tensor Cores

A tensor core performs multiplies an MxN matrice A by an NxP matrix B and accumulates the result in an MxP matrix C: $C_{n+1} = A_n B_n + C_n$ normally starting with C$_0$=0.

Each step is generally done in a single clock cycle and it is used to multiply larger matrices that have been partitioned into tiles, multiplying and accumulating a row by a column of tiles (see section 2.1 of [3]).

Fig.4 shows a possible architecture for a tensor core. By changing the processing elements (PEs, Fig.5) we can have an ordinary tensor core based on MACs or one based on partial multiplications.

We'll discuss the operations of a MAC based tensor core first that use PEs in Fig.5a. Note that the PEs do not need the Sa and Sb inputs in this case.

First the Init signal is raised and clears all the accumulators. Then each PE receives at its input a row of the matrix A and a column of the matrix B every clock cycle. The dot product between these two vectors is calculated and then accumulated. The result appears on the outputs O.

We now come to the partial multiplication version, based on the PE in Fig.5b. Here the Init signal does not clear the accumulator but initialise it to value Sa$_i$ + Sb$_j$.

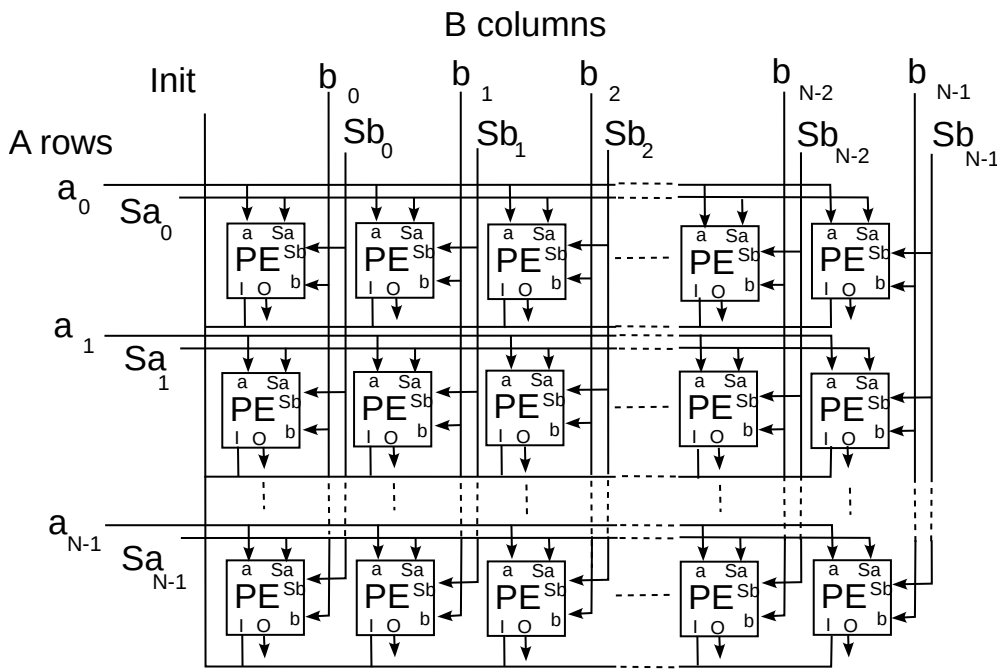

Figure 4A tensor core.

Note that, in this context, where the tensor core is multiplying a row and a column of tiles, $Sa_i$ and $Sb_j$ are the calculated from the ith row and jth column of the larger matrices (the ones that are being tiled) being multiplied.

After that, each PE receives a row and a column of B, performs a partial multiplication of all the elements of each vector that are then added together. We could call this a partial dot product which is then accumulated.

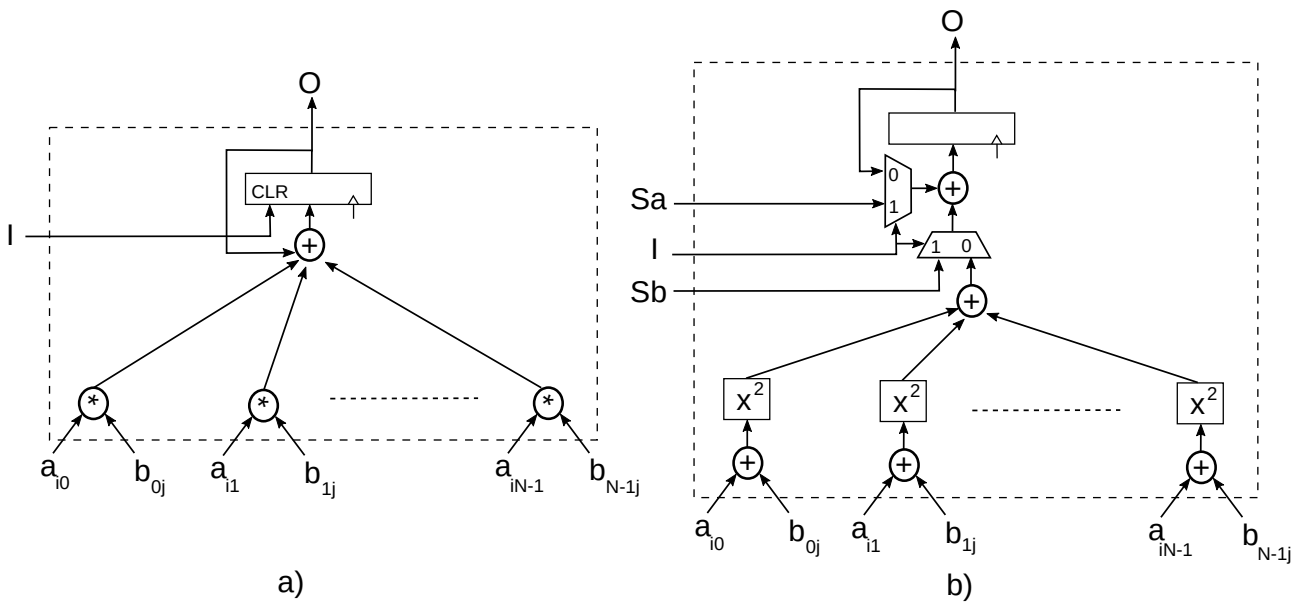

Figure 5PE for a MAC based tensor core a). For one based on partial multiplication.

Again, the result appears on the outputs O and it needs to be the corrected with single right shift when we are done.

# 4. Linear Transforms

We now examine the case of linear transforms, such as sine or cosine transforms:

$$X_k = \sum_{i=0}^{N-1} w_{ki} x_i \quad \forall k=0\ldots N-1 \quad (7)$$

We will concern ourselves with real values $x_i$ at first. The transform is defined by the coefficients $w_{ki}$, essentially a matrix/vector multiplication.

The architecture in Fig.6a shows a possible implementation of (7) using multipliers.

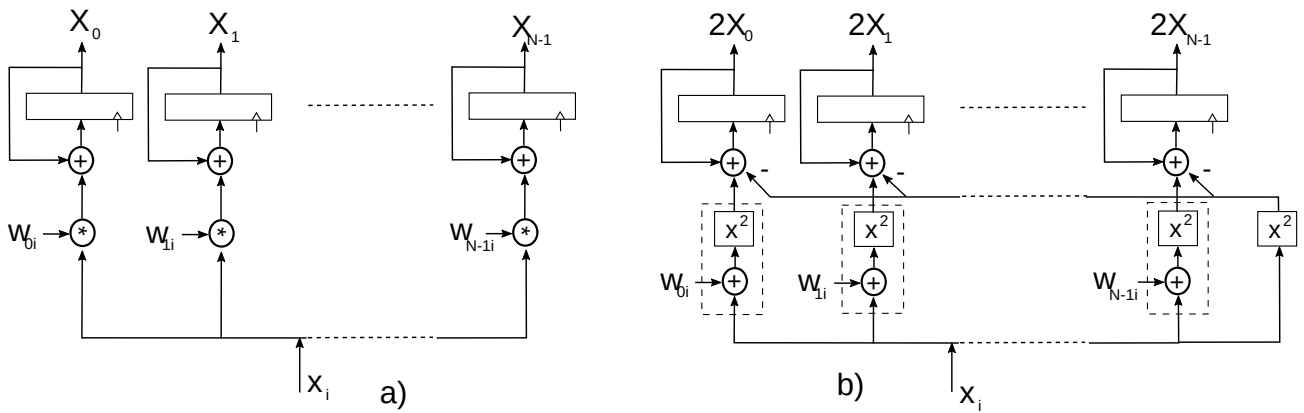

Figure 6: Linear transform with multipliers a) and squares b).

The registers $X_0, \ldots, X_{N-1}$ are initialized to zero. Then the values $x_0, \ldots, x_{N-1}$ are input one per clock cycle where they are simultaneously multiplied by the coefficients $w_{0i}, \ldots, w_{N-1i}$. After N clock cycles the registers $X_0, \ldots, X_{N-1}$ contain the result of the transform.

As before, we can use (1) to replace the multiplication $w_{ki}x_i$:

$$X_k = \sum_{i=0}^{N-1} w_{ki} x_i = \frac{1}{2} \sum_{i=0}^{N-1} \left( (w_{ki}+x_i)^2 - w_{ki}^2 - x_i^2 \right) = \frac{1}{2} \sum_{i=0}^{N-1} (w_{ki}+x_i)^2 - \frac{1}{2} \sum_{i=0}^{N-1} x_i^2 + \frac{1}{2} Sw_k \quad (8)$$

with:

$$Sw_k = -\sum_{i=0}^{N-1} w_{ki}^2 \quad (9)$$

From (8), the term $x_i^2$ is common and can be calculated once and subtracted from all the terms.

Unfortunately (8), unlike the case of matrix multiplication, does not offer any computational advantage as the cost of calculating $Sw_k$ is the same as the sum of partial multiplications.

However, often the coefficients $w_{ki}$ are constants and so the $Sw_k$ values can be pre-calculated. Even if the $w_{ki}$ coefficients are not constant, we don't generally perform a single transform but multiple one with the same coefficients. Therefore the cost of computing (9), in this case, can be seen as a single upfront cost to be considered in the context of multiple subsequent transformations.

This is more restrictive than the case of matrix multiplication but we can nevertheless proceed on this basis and assume the $Sw_k$ values are available. In this case we can use the architecture in the Fig.6b, with the partial multiplications encircled by the dashed lines.

In this architecture, the registers $X_0, …, X_{N-1}$ are initialized with the values $Sw_0,……,Sw_{N-1}$. We then input the values $x_0 ,……, x_{N-1}$ one per clock cycle. Each value $x_i$ is then added to the coefficients $w_{0i} ,……, w_{N-1 i}$ and then squared. The value $x_i^2$ is then subtracted from each result and then accumulated in the corresponding $X_k$ register. After N clock cycles the registers $X_0, …, X_{N-1}$ contain the result of the transform scaled, as usual, by a factor of 2.

It's clear that this architecture, still taking N cycles to complete has now N+1 squares instead of multipliers.

Note that the above also works when the coefficients $w_{ki}$ are complex values (but the values $x_i$ are still real): we just need two instantiations of the architecture in Fig6b: one for the real and one for the imaginary components.

This can, for example, cover the case of Discrete Fourier Transform of real values vectors.

We will discuss later the case where both the coefficients $w_{ki}$ and the values $x_i$ are both complex values.

# 5. Convolutions

Moving now to convolutions/correlations. We won't make a distinction between the two here because the implementation mechanism is essentially the same: we have a kernel of N weights $w_i$ sliding over real valued samples $x_i$ producing a single output at each step:

$$y_k = \sum_{i=0}^{N-1} w_i x_{i+k} \quad (10)$$

Fig.7a shows the classical implementation of (10). In this case the input samples moving through the registers are simultaneously multiplied by the kernel weights and added up to form the output.

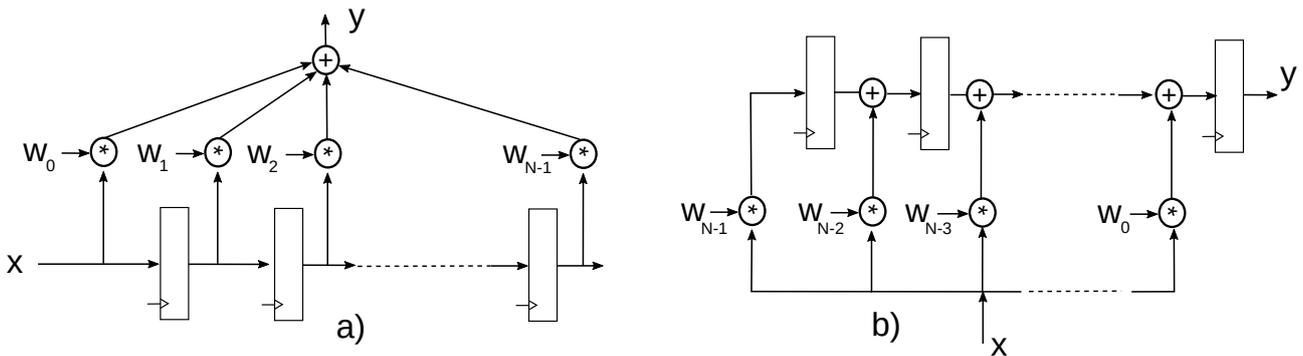

Figure 7: Convolution architectures.

Fig.7b shows an alternative, less common, implementation of (10). In this case each input sample is simultaneously multiplied by the kernel weights and added to separate registers until the final result is shifted out.

We now consider each multiplication $w_i x$ and replace it with (1):

$$w_i x = \frac{1}{2}((w_i+x)^2 - x^2 - w_i^2), \quad Sw = -\sum_{i=0}^{N-1} w_i^2 \quad (11)$$

We can replace the multipliers in Fig7.b and arrive at the architecture in the figure below.

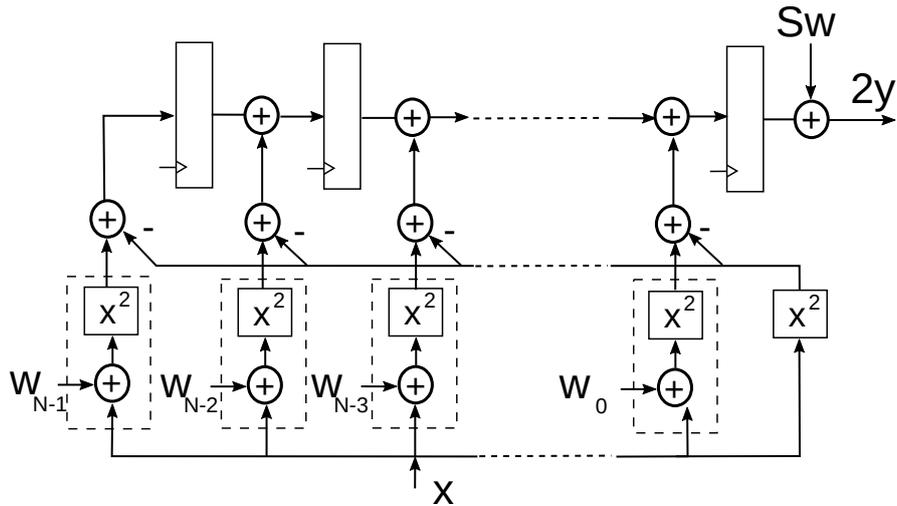

Figure 8: Square based convolution architecture.

Note that the term $x^2$ is common and it can be calculated once and subtracted from all the other terms. We would also need to subtract the terms $w_i^2$ from each register but as they get all propagated through we can subtract them all at once at the end.

Normally the kernel weights can be considered constants or, at the very least, changed infrequently and so the sum of their squares can be pre-calculated. In any case, convolutions are normally applied to thousands of samples and so the cost of calculating Sw can be distributed over multiple operations.

Again, we have effectively substituted N multipliers with N + 1 square operations. Note that the output needs to be scaled by a factor of 2.

The architectures in Fig.7a), Fig.7b) and Fig.8 can be used to compute FIR filters. For IIR filters we can apply the same principles.

## 5.1. 2D Convolutions

We'll briefly look at 2D convolutions. In this case we have a MxN sample kernel $w_{ij}$ sliding over samples $x_{h,k}$:

$$y_{hk} = \sum_{i=0}^{N-1} \sum_{j=0}^{M-1} w_{ij} x_{i+k, j+k} \quad (12)$$

As usual, we replace (1) in the product above:

$$y_{hk} = \sum_{i=0}^{N-1} \sum_{j=0}^{M-1} w_{ij} x_{i+k, j+k} = \frac{1}{2} \sum_{i=0}^{N-1} \sum_{j=0}^{M-1} \left( (w_{ij} + x_{i+k, j+k})^2 - x_{i+k, j+k}^2 - w_{ij}^2 \right) = \frac{1}{2} (Swx + Sx + Sw) \quad (13)$$

With:

$$Swx = \sum_{i=0}^{N-1} \sum_{j=0}^{M-1} (w_{ij} + x_{i+k, j+k})^2, \quad Swx = -\sum_{i=0}^{N-1} \sum_{j=0}^{M-1} x_{i+k, j+k}^2, \quad Sw = -\sum_{i=0}^{N-1} \sum_{j=0}^{M-1} w_{ij}^2 \quad (14)$$

Normally we can consider the weights $w_{ij}$ constant. So Sw can be pre-calculated or, if not, we normally perform the same convolution over many samples, diluting its cost.

We also note that, the same sample $x_{hk}$ is covered by multiple instances of the same kernel and so its corresponding $x_{hk}^2$ can be calculated once and shared amongst multiple kernel partial dot products. This is similar to what happens in the architecture in Fig.8 where it is calculated once and subtracted at multiple points.

## 6. Complex Matrix Multiplication

Let us now look at an MxN matrix X, an NxP matrix Y and their product, the MxP matrix Z, with $a_{hk}+jb_{hk}$ an element of X, $c_{hk}+js_{hk}$ an element of Y and $z_{hk}$ an element of Z, all complex numbers:

$$z_{hk} = \sum_{i=0}^{N-1} (a_{hi}+jb_{hi})(c_{ik}+js_{ik}) \quad \forall h=0 \ldots M-1, k=0 \ldots P-1 \quad (15)$$

Then:

$$z_{hk} = \sum_{i=0}^{N-1} (a_{hi}+jb_{hi})(c_{ik}+js_{ik}) = \sum_{i=0}^{N-1} (a_{hi}c_{ik} - b_{hi}s_{ik}) + j(b_{hi}c_{ik} + a_{hi}s_{ik}) \quad (16)$$

Taking the real part of $z_{hk}$, replacing $a_{hi}c_{ik}$ with (1) and $-b_{hi}s_{ik}$ with (2):

$$\Re(z_{hk}) = \sum_{i=0}^{N-1} (a_{hi}c_{ik} - b_{hi}s_{ik}) = \frac{1}{2}\sum_{i=0}^{N-1} ((a_{hi}+c_{ik})^2 - a_{hi}^2 - c_{ik}^2 + (b_{hi}-s_{ik})^2 - b_{hi}^2 - s_{ik}^2) =$$

$$= \frac{1}{2}\sum_{i=0}^{N-1} ((a_{hi}+c_{ik})^2 + (b_{hi}-s_{ik})^2) + \frac{1}{2}Sx_h + \frac{1}{2}Sy_k \quad (17)$$

With:

$$Sx_h = -\sum_{i=0}^{N-1}(a_{hi}^2+b_{hi}^2), \quad Sy_k = -\sum_{i=0}^{N-1}(c_{ik}^2+s_{ik}^2) \quad (18)$$

Note that if the elements of either the matrix X or Y are unit complex numbers (like, for example, in the case of the DFT matrix [4]), then the equations (18) simplify to -N.

Then we take the imaginary part of $z_{hk}$ and replace both $b_{hi}c_{ik}$ and $a_{ih}s_{ik}$ with (1):

$$\Im(z_{hk}) = \sum_{i=0}^{N-1}(b_{hi}c_{ik}+a_{hi}s_{ik}) = \frac{1}{2}\sum_{i=0}^{N-1}((b_{hi}+c_{ik})^2 - b_{hi}^2 - c_{ik}^2 + (a_{hi}+s_{ik})^2 - a_{hi}^2 - s_{ik}^2) =$$

$$= \frac{1}{2}\sum_{i=0}^{N-1}((b_{hi}+c_{ik})^2 + (a_{hi}+s_{ik})^2) + \frac{1}{2}Sx_h + \frac{1}{2}Sy_k \quad (19)$$

It is clear from (17) and (19) that the situation is analogous to section 3 and $Sx_h$ and $Sy_k$ can be re-used.

In order to calculate the product of the matrices X and Y, we need M*N*P complex multiplications. If we calculate it with (17) and (19), we need 4*M*N*P squares plus 2*M*N squares + 2*N*P squares. The ratio between the number of complex multiplications and the number of squares is:

$$\frac{4MNP+2MN+2NP}{MNP} = 4 + \frac{2}{P} + \frac{2}{M} \quad (20)$$

This ratio tends to 4 quickly as M and P increase. We effectively need 4 square operation for each complex multiplication for many practical matrices sizes.

## 6.1. Hardware Implementation

The discussion here is essentially the same as in section 3.1, except that the partial multiplication is now a complex partial multiplication. The complex partial multiplication between a+jb and c+js is defined, from (17), for the real part, as:

$$(a+c)^2 + (b-s)^2 \quad (21)$$

and, from (19), for the imaginary part, as:

$$(b+c)^2 + (a+s)^2 \quad (22)$$

Fig.9a shows a possible implementation of a complex partial multiplication (CPM). Fig.9b shows a complex multiplier implemented with three real multipliers, for comparison.

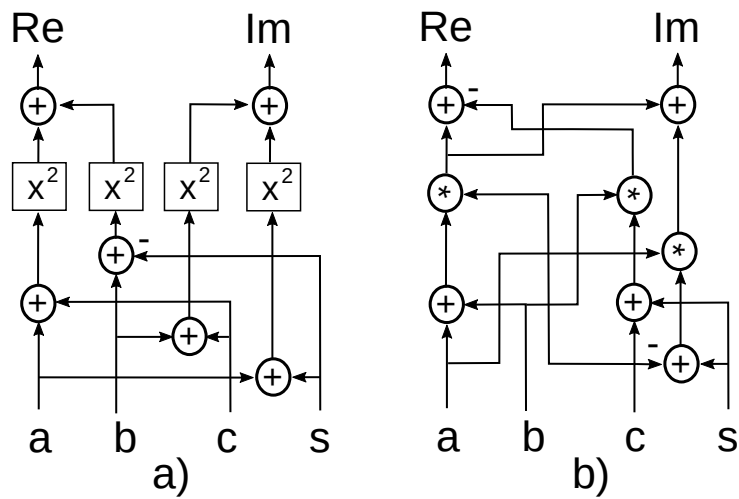

Figure 9: Partial complex multiplier a). Complex multiplier b).

And, as discussed in section 3.1, it can replace a complex multiplier in any MACs when computing (17) and (19), as long as we correct the final result by adding $(Sx_h + Sy_k)(1+j)$. We also need to divide the final result by 2. Thus the same discussions regarding various architectures described in section 3.1 could be repeated here with small adaptations to CPMs.

## 7. Complex Linear Transforms

We now look at linear transforms again in the case where both the coefficients and the samples are complex values:

$$X_k + jY_k = \sum_{i=0}^{N-1} (c_{ki} + js_{ki})(x_i + jy_i) = \sum_{i=0}^{N-1} ((c_{ki}x_i - s_{ki}y_i) + j(c_{ki}y_i + s_{ki}x_i)) \quad \forall k = 0 \ldots N-1 \quad (23)$$

The transform is defined by the complex values $c_{ki} + js_{ki}$ and it could be, for example the Discrete Fourier Transform DFT.

We now take the real part of (23), replacing $c_{ki}x_i$ with (1) and $-s_{ki}y_i$ with (2):

$$X_k = \sum_{i=0}^{N-1} (c_{ki}x_i - s_{ki}y_i) = \frac{1}{2} \sum_{i=0}^{N-1} ((c_{ki} + x_i)^2 - c_{ki}^2 - x_i^2 + (s_{ki} - y_i)^2 - s_{ki}^2 - y_i^2) =$$

$$= \frac{1}{2}\sum_{i=0}^{N-1}\left((c_{ki}+x_i)^2+(s_{ki}-y_i)^2\right)+\frac{1}{2}Sxy+\frac{1}{2}S_k \quad (24)$$

With:

$$Sxy=-\sum_{i=0}^{N-1}(x_i^2+y_i^2) \quad S_k=-\sum_{i=0}^{N-1}(c_{ki}^2+s_{ki}^2) \quad (25)$$

Again, as in section 6, if the elements of the transform are unit complex numbers, then $S_k$ simplifies to -N. Next we take the imaginary part of (23) and replace both $c_{ik}y_i$ and $s_{ik}x_i$ with (1):

$$Y_k=\sum_{i=0}^{N-1}(c_{ki}y_i+s_{ki}x_i)=\frac{1}{2}\sum_{i=0}^{N-1}\left((c_{ki}+y_i)^2-c_{ki}^2-y_i^2+(s_{ki}+x_i)^2-s_{ki}^2-x_i^2\right)=$$

$$=\frac{1}{2}\sum_{i=0}^{N-1}\left((c_{ki}+y_i)^2+(s_{ki}+x_i)^2\right)+\frac{1}{2}Sxy+\frac{1}{2}S_k \quad (26)$$

As far as hardware implementation is concerned, we can make the exact same considerations as in section 4 and create the same architecture as in Fig.6b, see Fig.10.

The registers now hold complex values and the adders work on complex numbers. Instead of partial multiplications, we now have complex partial multiplications (CPMs), as previously defined in section 6.1, that we can recognise as part of (24) and (26).

According to (24) and (26), the term $(x_i^2+y_i^2)(1+j)$ is common and can be calculated once and subtracted from all the terms.

The architecture works in the same way as the one in Fig.6b. At the start the accumulating registers are initialised with $S_0(1+j)$, $S_1(1+j)$, …., $S_{N-1}(1+j)$.

Every clock cycle a sample $x_0+jy_{0i}$, $x_1+jy_1$, … $x_{N-1}+jy_{N-1}$ is input and simultaneously partially multiplied by the CPMs to $c_{0i}+js_{0i}$, $c_{1i}+js_{1i}$, … $c_{N-1i}+js_{N-1i}$. The term $(xi^2+yi^2)(1+j)$ is also subtracted to all of them and the result accumulated in the registers.

After N clock cycles the registers will contain two times the result of the transform.

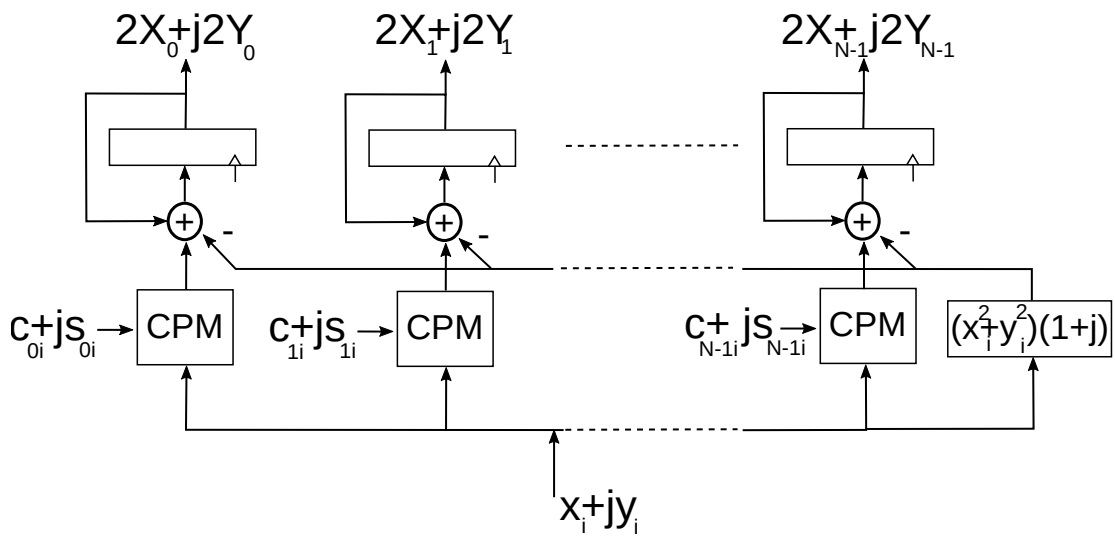

Figure 10: Linear transform for complex numbers using squares.

Note, as usual, that after the computation is over, the complex registers need to be scaled by a factor of 2.

# 8. Complex Convolutions

We look at convolutions/correlations again with a kernel of N complex weights $c_i+js_i$ sliding over complex valued samples $x_i+jy_i$ producing a single complex output $z_k$ at each step:

$$z_k = \sum_{i=0}^{N-1} (c_i+js_i)(x_{i+k}+jy_{i+k}) = \sum_{i=0}^{N-1} ((c_i x_{i+k} - s_i y_{i+k}) + j(c_i y_{i+k} + s_i x_{i+k})) \quad (27)$$

The architecture in Fig.7b is suitable for complex weights and samples. So we can modify it taking into account the real and imaginary part of the product in (27).

For the real part, we $c_i x_{i+k}$ with (1) and $-s_i y_{i+k}$ with (2):

$$c_i x_{i+k} - s_i y_{i+k} = \frac{1}{2}((c_i+x_{i+k})^2 - x_{i+k}^2 - c_i^2 + (s_i - y_{i+k})^2 - y_{i+k}^2 - s_i^2) =$$

$$= \frac{1}{2}((c_i+x_{i+k})^2 + (s_i - y_{i+k})^2 - x_{i+k}^2 - y_{i+k}^2 - c_i^2 - s_i^2) \quad (28)$$

For the imaginary part we $c_i y_{i+k}$ and $s_i x_{i+k}$ with (1):

$$c_i y_{i+k} + s_i x_{i+k} = \frac{1}{2}((s_i+x_{i+k})^2 - x_{i+k}^2 - s_i^2 + (c_i+y_{i+k})^2 - y_{i+k}^2 - c_i^2) =$$

$$= \frac{1}{2}((s_i+x_{i+k})^2 + (c_i+y_{i+k})^2 - x_{i+k}^2 - y_{i+k}^2 - c_i^2 - s_i^2) \quad (29)$$

With:

$$Sw = -\sum_{i=0}^{N-1} (c_i^2 + s_i^2) \quad (30)$$

We recognise the complex partial multiplication as defined in section 6.1 in (28) and (29) which will then replace the partial multiplications in the architecture in Fig.8 in order to arrive at the architecture in Fig.11. If the weights are unit complex numbers, then $S_w$ simplifies to -N.

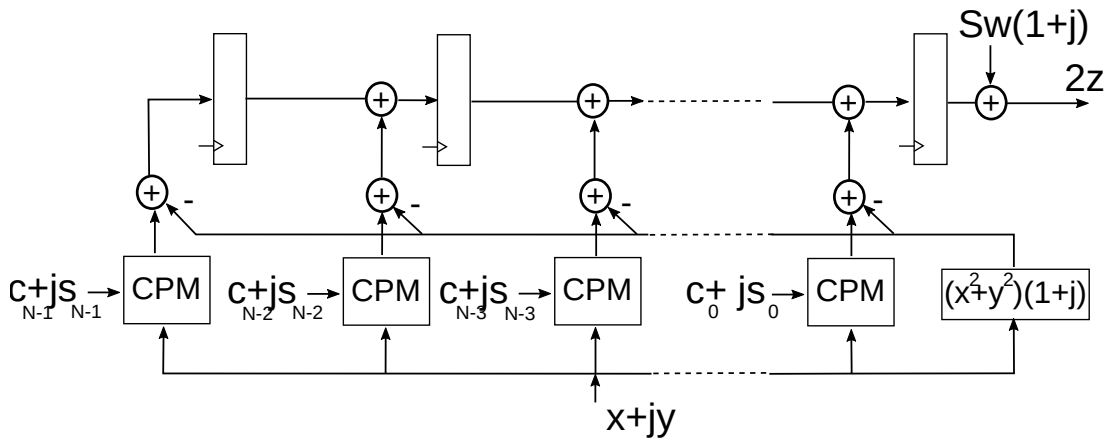

Figure 11  Square based convolution architecture with complex weights and values.
Again, the architecture outputs 2 times the correct output.

## 9. Complex Matrix Multiplication with Three Squares

We now go back to section 6 and rewrite the complex multiplication in (15) with 3 real multiplications:

$$z_{hk} = \sum_{i=0}^{N-1} (a_{hi} + jb_{hi})(c_{ik} + js_{ik}) = \sum_{i=0}^{N-1} ((c_{ik}(a_{hi}+b_{hi}) - b_{hi}(c_{ik}+s_{ik})) + j(c_{ik}(a_{hi}+b_{hi}) + a_{hi}(s_{ik}-c_{ik}))) \quad (31)$$

We then replace both products in the real part of (31) with (1):

$$\Re(z_{hk}) = \sum_{i=0}^{N-1} (c_{ik}(a_{hi}+b_{hi}) - b_{hi}(c_{ik}+s_{ik})) =$$

$$= \frac{1}{2} \sum_{i=0}^{N-1} ((c_{ik}+a_{hi}+b_{hi})^2 - (b_{hi}+c_{ik}+s_{ik})^2 - (a_{hi}+b_{hi})^2 + b_{hi}^2 - c_{ik}^2 + (c_{ik}+s_{ik})^2) =$$

$$= \frac{1}{2} \sum_{i=0}^{N-1} ((c_{ik}+a_{hi}+b_{hi})^2 - (b_{hi}+c_{ik}+s_{ik})^2) + \frac{1}{2} Sab_h + \frac{1}{2} Scs_k \quad (32)$$

With:

$$Sab_h = \sum_{i=0}^{N-1} (-(a_{hi}+b_{hi})^2 + b_{hi}^2), \quad Scs_k = \sum_{i=0}^{N-1} (-c_{ik}^2 + (c_{ik}+s_{ik})^2) \quad (33)$$

Now we replace both products in the imaginary part of (31) with (1):

$$\Im(z_{hk}) = \sum_{i=0}^{N-1} (c_{ik}(a_{hi}+b_{hi}) + a_{hi}(s_{ik}-c_{ik})) =$$

$$= \frac{1}{2} \sum_{i=0}^{N-1} ((c_{ik}+a_{hi}+b_{hi})^2 + (a_{hi}+s_{ik}-c_{ik})^2 - (a_{hi}+b_{hi})^2 - a_{hi}^2 - c_{ik}^2 - (s_{ik}-c_{ik})^2) =$$

$$= \frac{1}{2} \sum_{i=0}^{N-1} ((c_{ik}+a_{hi}+b_{hi})^2 + (a_{hi}+s_{ik}-c_{ik})^2) + \frac{1}{2} Sba_h + \frac{1}{2} Ssc_k \quad (34)$$

With:

$$Sba_h = \sum_{i=0}^{N-1} (-(a_{hi}+b_{hi})^2 - a_{hi}^2), \quad Ssc_k = \sum_{i=0}^{N-1} (-c_{ik}^2 - (s_{ik}-c_{ik})^2) \quad (35)$$

Note that (32) and (34) have the term $(c_{ik}+a_{hi}+b_{hi})^2$ in common. This means that there are only 3 square operations that depend on both indexes h and k and thus we need a total 3*M*N*P squares to calculate them all.

As for the terms $Sab_h$, $Sba_h$, $Scs_k$ and $Ssc_k$, they depend on the indexes h and k separately, they can be re-used, and thus they can be calculated with a total of 3*MN+3NP square operations.

We can now calculate the ratio between the total number of squares required to calculate (31) and the total number of complex multiplications:

$$\frac{3\,MNP + 3\,MN + 3\,NP}{MNP} = 3 + \frac{3}{P} + \frac{3}{M} \quad (36)$$

This ratio tends to 3 quickly as M and P increase. We effectively need 3 square operation for each complex multiplication for many practical matrices sizes.

## 9.1. Hardware Implementation

The discussion here is essentially the same as in section 6.1, except that the complex partial multiplication will now use 3 square operations. The new complex partial multiplication between a+jb and c+js is defined, from (32), for the real part, as:

$$(c+a+b)^2 - (b+c+s)^2 \quad (37)$$

and, from (34), for the imaginary part, as:

$$(c+a+b)^2 + (a+s-c)^2 \quad (38)$$

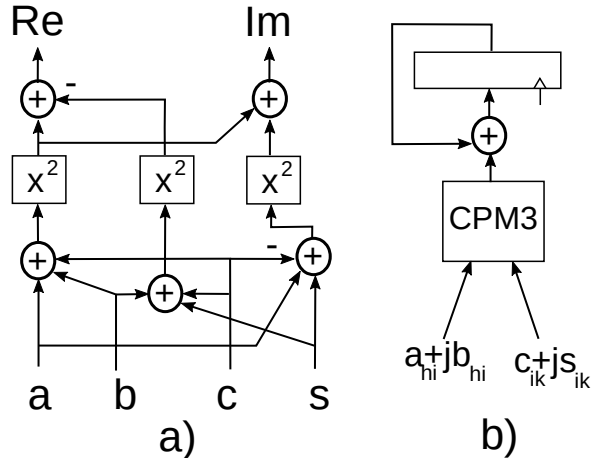

Figure 12: CPM3 with 3 squares a). Complex CPM/3 accumulator.

Fig.12a shows a possible implementation of a complex partial multiplier with 3 square operations (CPM3 to distinguish it from the previously defined CPM).

And, as discussed in section 6.1, it can replace a complex multiplier in any MACs when computing (32) and (34), as long as we correct the final result by adding $(Sab_h + Scs_k) + j(Sba_h + Ssc_k)$. So, Fig.12b shows a complex partial multiply accumulator. We can initialise it with $(Sab_h + Scs_k) + j(Sba_h + Ssc_k)$ and then simply input $(a_{hi} + jb_{hi})$ and $(c_{ik} + js_{ik})$ at each cycle to calculate (31). When we have done, the result will be $2z_{ij}$.

Thus the same discussions regarding various architectures described in section 3.1 could be repeated here with small adaptations to CPM3s.

## 10. Revisiting Complex Linear Transforms

We now look at linear transforms again from section 7 and rewrite the complex product in (23) with 3 real multiplications:

$$X_k + jY_k = \sum_{i=0}^{N-1} \left( \left( c_{ki}(x_i + y_i) - y_i(c_{ki} + s_{ki}) \right) + j\left( c_{ki}(x_i + y_i) + x_i(s_{ki} - c_{ki}) \right) \right) \quad (39)$$

The transform is defined by the complex values $c_{ki} + js_{ki}$. We take the real part of (39), replacing the products with (1):

$$X_k = \frac{1}{2} \sum_{i=0}^{N-1} \left( (c_{ki} + x_i + y_i)^2 - (y_i + c_{ki} + s_{ki})^2 - (x_i + y_i)^2 + y_i^2 - c_{ki}^2 + (c_{ki} + s_{ki})^2 \right) =$$

$$= \frac{1}{2} \sum_{i=0}^{N-1} \left( (c_{ki}+x_i+y_i)^2 - (y_i+c_{ki}+s_{ki})^2 \right) + \frac{1}{2} Sxy + \frac{1}{2} Sx_k \quad (40)$$

With:

$$Sxy = \sum_{i=0}^{N-1} \left( -(x_i+y_i)^2 + y_i^2 \right), \quad Sx_k = \sum_{i=0}^{N-1} \left( -c_{ki}^2 + (c_{ki}+s_{ki})^2 \right) \quad (41)$$

We take the imaginary part of (39), replacing the products with (1):

$$Y_k = = \frac{1}{2} \sum_{i=0}^{N-1} \left( (c_{ki}+x_i+y_i)^2 + (x_i+s_{ki}-c_{ki})^2 - (x_i+y_i)^2 - x_i^2 - c_{ki}^2 - (s_{ki}-c_{ki})^2 \right) =$$

$$= \frac{1}{2} \sum_{i=0}^{N-1} \left( (c_{ki}+x_i+y_i)^2 + (x_i+s_{ki}-c_{ki})^2 \right) + \frac{1}{2} Syx + \frac{1}{2} Sy_k \quad (42)$$

With:

$$Syx = \sum_{i=0}^{N-1} \left( -(x_i+y_i)^2 - x_i^2 \right), \quad Sy_k = \sum_{i=0}^{N-1} \left( -c_{ki}^2 + (s_{ki}-c_{ki})^2 \right) \quad (43)$$

We can use the same architecture in Fig.10, shown in Fig.13, with the following differences:

- Use of CPM3 (recognisable from (40) and (42)) instead of CPM
- Each registers k will be initialised to the complex values Sxk+jSyk
- The common term to be subtracted from all the others is now $(-(x_i+y_i)^2+y_i^2)+j(-(x_i+y_i)^2-x_i^2)$

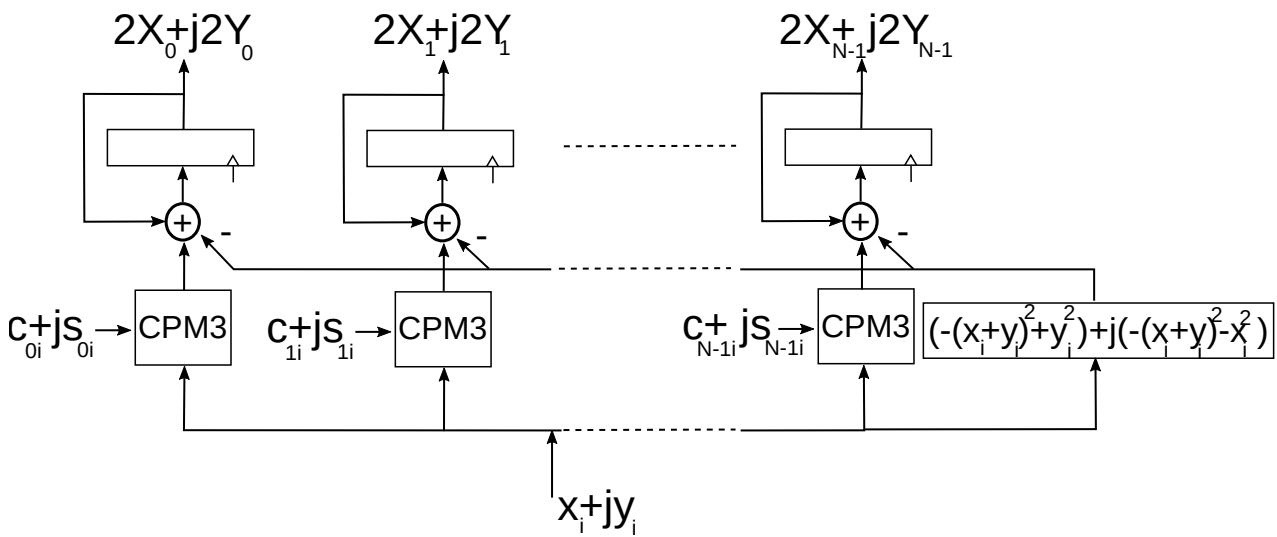

Figure 13: Complex linear transform with CPM3.

Again, the architecture outputs 2 times the correct output.

## 11. Revisiting Complex Convolutions

We now look at linear transforms again from section 8 and rewrite the complex product in (27) with 3 real multiplications:

$$z_k = \sum_{i=0}^{N-1} \left( (c_i(x_{i+k}+y_{i+k}) - y_{i+k}(c_i+s_i)) + j(c_i(x_{i+k}+y_{i+k}) + x_{i+k}(s_i-c_i)) \right) \quad (44)$$

For the real part we replace the products with (1):

$$c_i(x_{i+k}+y_{i+k}) - y_{i+k}(c_i+s_i) =$$

$$= \frac{1}{2}\left( (c_i+x_{i+k}+y_{i+k})^2 - (y_{i+k}+c_i+s_i)^2 - (x_{i+k}+y_{i+k})^2 + y_{i+k}^2 - c_i^2 + (c_i+s_i)^2 \right) \quad (45)$$

For the imaginary part part we replace the products with (1):

$$c_i(x_{i+k}+y_{i+k}) + x_{i+k}(s_i-c_i) =$$

$$= \frac{1}{2}\left( (c_i+x_{i+k}+y_{i+k})^2 + (x_{i+k}+s_i-c_i)^2 - (x_{i+k}+y_{i+k})^2 - x_{i+k}^2 - c_i^2 - (s_i-c_i)^2 \right) \quad (46)$$

With:

$$Sw = \sum_{i=0}^{N-1} \left( (-c_i^2 + (c_i+s_i)^2) + j(-c_i^2 - (s_i-c_i)^2) \right) \quad (47)$$

We can use the same architecture in Fig.11, shown in Fig.14, with the following differences:

- Use of CPM3 (recognisable from (40) and (42)) instead of CPM
- The result is adjusted by adding Sw
- The common term to be subtracted from all the others is now $(-(x+y)^2+y^2)+j(-(x+y)^2-x^2)$

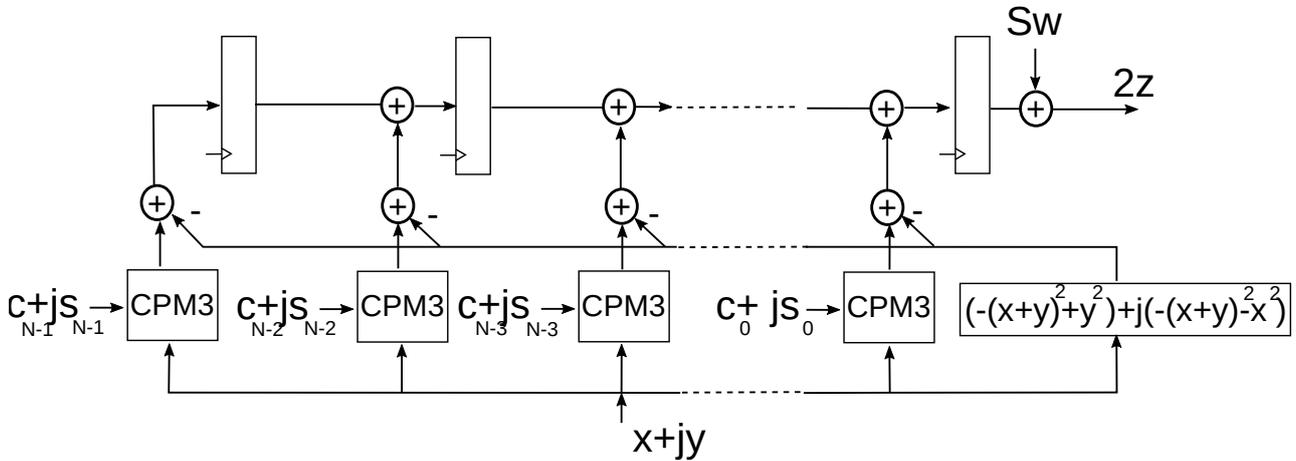

Figure 14: Complex convolutions with CPM3.

Needless to say, the architecture outputs 2 times the correct output.

## 12. Conclusion

A simple method to asymptotically replace one multiplier with one square operation when performing matrix multiplication was described.

The method was extended to complex matrix multiplication where it was shown to asymptotically require 4 and then 3 square operations for complex multiplications.

Also, the same results were shown for convolutions and linear transforms.

Finally, a variety of architectures to take advantage of this have been proposed from systolic arrays to tensor cores and convolution engines. Since a square circuit requires roughly half the gate count of a multiplier, this results in large savings in area and power in digital designs.